\documentclass[prl,aps,twocolumn,showpacs,a4paper]{revtex4}
\usepackage{bm}
\usepackage[dvips]{graphicx}
\usepackage{amsmath}
\usepackage{subfigure}
\usepackage{verbatim}
\begin{document}
\newcommand{\la}{\label}
\newcommand{\be}{\begin{eqnarray}}
\newcommand{\ee}{\end{eqnarray}}
\newcommand{\der}{\partial}
\newcommand{\w}{\tilde}
\newcommand{\sgn}{\,{\mathrm{sgn}}}
\newcommand{\tr}{\,{\mathrm{tr}}\,}
\newcommand{\re}{{\mathrm{Re}}\,}
\newcommand{\im}{{\mathrm{Im}}\,}
\newcommand{\e}{_{\mathrm{eq}}}

\title{Fluid Velocity Fluctuations in a Suspension of Swimming Protists}

\author{Ilia Rushkin$^{1,2}$, Vasily Kantsler$^{1}$ and Raymond E. Goldstein$^{1}$}
\affiliation{$^{1}$Department of Applied Mathematics and Theoretical
Physics, Centre for Mathematical Sciences, University of Cambridge, Wilberforce Raod, Cambridge CB3 0WA, UK\\
$^{2}$School of Mathematical Sciences, University of Nottingham, Nottingham NG7
2RD, UK}

\date{\today}

\begin{abstract}
In dilute suspensions of swimming microorganisms the local fluid velocity is a
random superposition of the flow fields set up by the individual organisms, which in turn
have multipole contributions decaying as inverse powers of distance from the 
organism. Here we show that the conditions under which the central limit theorem guarantees 
a Gaussian probability distribution function of velocities
are satisfied when the leading force singularity is a Stokeslet, but are 
not when it is any higher multipole. These results are confirmed by numerical studies 
and by experiments on suspensions 
of the alga {\it Volvox carteri}, which show that deviations from Gaussianity arise from near-field
effects. 
\end{abstract}


\maketitle

A key feature of the inertialess world inhabited by microscopic 
organisms is the very long-range flow fields they create as they swim.  
For neutrally-buoyant, self-propelled organisms the far-field behavior
of the velocity is that of the force dipole (stresslet) created by the opposed actions of
their flagella and cell body on the fluid.  Theories incorporating such fields 
in the fluid stress tensor \cite{Ramaswamy}, and  
simulations of suspensions of dipolar organisms \cite{HernandezOrtiz} have shown
the formation of large coherent structures that 
are highly suggestive of those seen in experiments on the bacterium 
{\it B. subtilis} \cite{Dombrowski}.
The suggestion \cite{Dombrowski} that hydrodynamic interactions
underlie these vortices and jets was made by analogy with the appearance of similar patterns in 
suspensions of sedimenting particles \cite{Segre}, although interactions 
in the latter are due to the force \textit{monopole} (Stokeslet) 
fields arising from the density mismatch between the particles and fluid.  
Although Stokeslet and stresslet fields have different topologies, it is 
striking that the two systems display similar coherence.

The relationship between suspensions of microorganisms and of sedimenting 
particles takes on new significance in light of measurements 
of velocity fields around freely-swimming organisms \cite{Drescher10}, which emphasized 
that the Stokeslet field dominates that of the stresslet 
beyond a length $\Lambda\sim Td/F_g$, where $d$ is the 
offset between the flagellar thrust 
$T$ and the body drag, and $F_g$ is the net gravitational force per organism.
$\Lambda$ can be surprisingly small when compared to the organism radius $R$:
while for the unicellular alga {\it Chlamydomonas reinhardtii} \cite{harris09} ($R\sim 5$ $\mu$m)  
$\Lambda\sim 30R$, for its multicellular descendant 
{\it Volvox carteri} \cite{Kirkbook} ($R\sim 200-400$ $\mu$m) there is the 
striking conclusion that $\Lambda\sim R$; the Stokeslet dominates 
the flow field.  It was therefore suggested \cite{Drescher10} that suspensions
of {\it Volvox} would be more 
similar to those of sedimenting particles than previously thought, the chief difference
being the component of the organism's motion from active swimming.
Hence there is fundamental interest in the question:
\textit{What are the statistics of fluid velocity fluctuations in a suspension of swimming microorganisms?}

\begin{figure*}[t]
\begin{center}
\includegraphics*[clip=true,width=1.9\columnwidth]{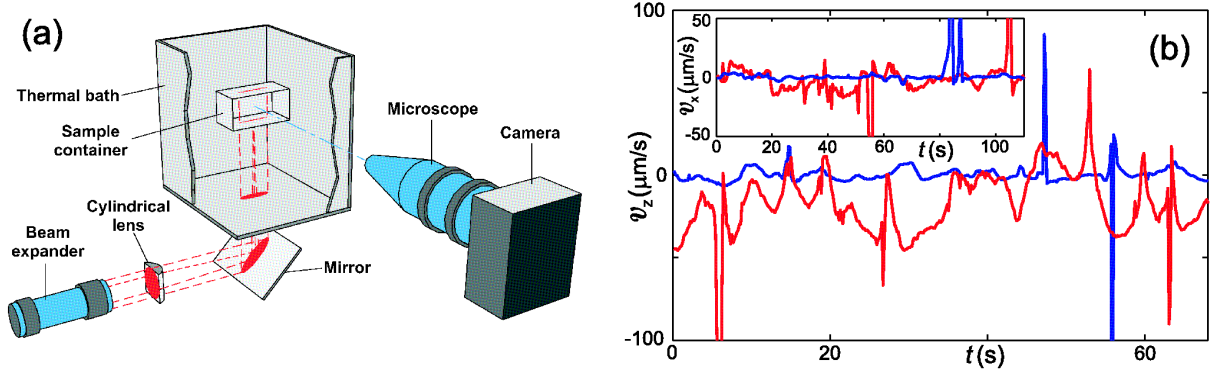}
\end{center}
\caption{\label{apparatus_example} (color online). Experimental setup and measured velocity fluctuations. 
(a) Schematic of the imaging and illumination system. (b) Experimental $z$-component 
(and $x$-component, inset) of the fluid 
velocity in a {\it Volvox} suspension ($N=300$ cm$^{-3}$) as a function of time
in the central PIV grid domain of the chamber. 
Red lines indicate velocities in SVM, while blue lines are with approximate 
density matching of the external fluid: SVM+3$\%$ v/v Percoll (Sigma).}
\label{figure1}
\end{figure*}

Here we present theory, experiments and simulations that 
elucidate a number of aspects of this question.   
We determine the condition on the leading force singularity 
of a swimmer in order that a random superposition of its
velocity field has a Gaussian probability distribution function
(PDF).  The condition admits the Stokeslet but excludes the stresslet and higher
multipoles, so the presence or absence of density matching has a
qualitative effect on the statistics. 
The velocity distribution functions found in experiment and simulation display 
clear non-Gaussian tails which we suggest arise from near-field effects \cite{Dunkel}.  
The large size of {\it Volvox} allows study of the scale of fluctuations as a function of
the number of organisms at fixed container size, complementary to the 
limiting procedure often adopted in sedimentation \cite{CaflischLuke}.  
Our result complement recent studies of the short-time PDFs of tracer particles in suspensions
of {\it Chlamydomonas} \cite{Leptos}, where non-Gaussianity was found, and to studies 
of fluctuations in bacterial baths \cite{Yodh,SokolovPNAS}.

Consider a suspension in a box of linear scale $L$, with $N$ swimmers of radius $R$.  If the volume 
fraction $\phi=4\pi R^3N/3L^3$ is
sufficiently small, the PDF of velocities 
reflects the statistics of a \textit{random} superposition of the flow fields 
around each swimmer. For a uniform spatial distribution of swimmers, averaging over their positions
is equivalent to integrating over space with the swimmer at the origin. If the velocity around a swimmer
decays as $\vert{\bf v}(r)\vert\sim A(\Omega)/r^n$, with $\Omega$ standing for 
angular variables, the probability
distribution $P(v)$ of velocities is
\begin{equation}
\la{eq1}
P(v)\propto L^{-3}\int_0^L\int_{D_{\Omega}}\delta\Bigl(v-\frac{A(\Omega)}{r^{n}}\Bigr)r^2 dr d\Omega ~,
\end{equation}
assuming a spherical container. The tail of the distribution can be determined from the behavior
of $P$ under the rescaling $r\to ar$.  Since $\delta(v-A/(ar)^n)=a^n\delta(va^n-A/r^n)$, and noting that
for large $v$ the argument of the $\delta$-function vanishes at small $r$, we deduce that
the integral does not depend on the upper limit $L/a$ (which can be taken to $\infty$), and hence
\begin{equation}
P(v)=a^{3+n}P(va^n) \to P(v)\propto\frac{1}{v^{1+3/n}}.
\label{scaling}
\end{equation}
The second moment of $P(v)$ is finite only if $n<3/2$, the case of a 
Stokeslet ($n=1$). This is the condition for validity of the central limit theorem; the
velocity field from a large number of independently placed Stokeslets 
is Gaussian. It will not be so for any higher integer singularity, such as stresslets ($n=2$) 
or source doublets (here termed `sourcelets') ($n=3$) \cite{BlakeChwang74}.  
If below a certain radius the decay law deviates from $v\propto r^{-n}$, 
the PDF shape (\ref{scaling}) will break down above the corresponding value of $v$.  Hence, 
deviations from Gaussianity provide a direct probe of the near-field velocity around the swimmers.

The spherical colonial alga {\it Volvox} is a remarkably useful system
for the study of many aspects of biological fluid dynamics 
\cite{Solari06,Short06,waltzing,pnas_phototaxis} because of its size, high symmetry, 
ease of growth, well-characterized biology, and the existence of a range of mutants.  
In our experiments, {\it Volvox carteri f. nagariensis} strain EVE were 
grown axenically in SVM \cite{kirk83} in a diurnal growth chamber set to a cycle of 16 hours 
artificial cool daylight ($\sim 4000$ lux) at $28^\circ$C and 8 h in the dark at $26^\circ$C. 
We used synchronized colonies from the first day of 
the $48$ hour life cycle to obtain the highest motility. A concentration 
$c=10-500$ cm$^{-3}$ (a volume fraction below $\phi=0.015$) of organisms was 
prepared in SVM, with added $2$ $\mu$m polystyrene seeding particles 
or 6 $\mathrm{\mu m}$ tracer particles (Polysciences) at a concentration
of $\sim 25\, \mathrm{ppm}$,
and placed into a glass container ($1\times 1\times 1$ cm). The container was 
placed a thermal bath (Fig. \ref{figure1}a) to eliminate convection \cite{RevSciInst}, and 
was illuminated with a thin laser sheet ($\lesssim 300\,\mathrm{\mu m}$) 
from a $100$ mW, $635$ nm laser (BWTEK).  Video was captured at frame rates of $0.4-5$ Hz by a CCD
camera (Pike F145B, Allied Vision Technologies) connected to a long-working distance microscope 
(Infinivar CFM-2/S, Infinity Photo-Optical).
The fluid velocity was measured by PIV (Dantec Dynamics),
typically producing a $63\times 63$ rectangular lattice of velocity vectors. Alternatively, we  
measured tracer and {\it Volvox} trajectories by custom (Matlab) PTV software.

\begin{figure*}[htp]
\begin{center}
\includegraphics[clip=true,width=2.0\columnwidth]{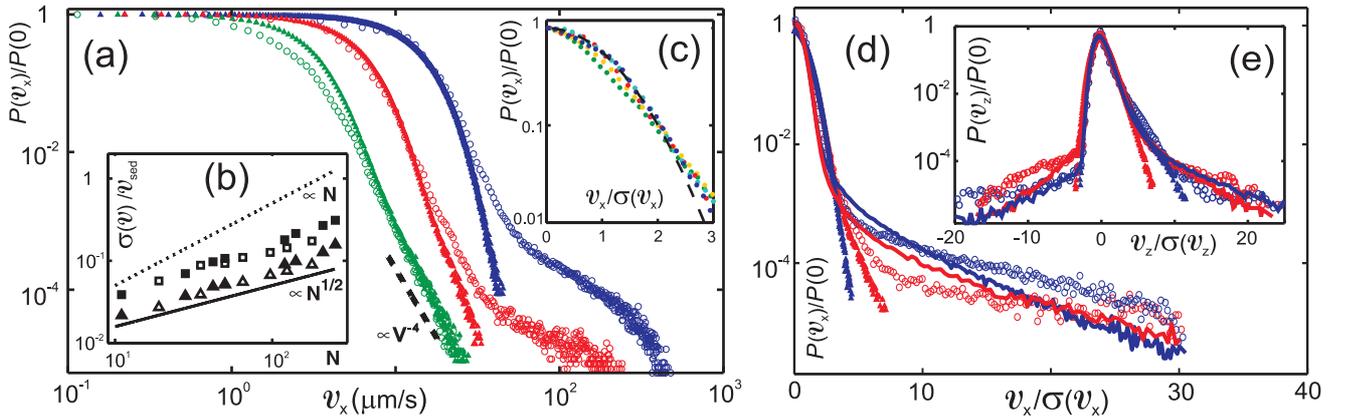}
\end{center}
\caption{(color online). Statistics of velocity fluctuations. 
Colored circles are experimental data for a suspension of {\it Volvox} with mean diameter 
$300$ $\mu$m. Colored triangles are corresponding numerical simulations 
excluding sourcelets; solid lines at right are for simulations including sourcelets ($\mu=4$). 
Individual colors indicate
different numbers of colonies in container: green ($11$), red ($42$),  
and blue ($210$). (a) PDF of fluctuations in horizontal velocity $v_x$. (b) Standard deviation of fluid velocity 
normalized by colony sedimentation speed $v_{\rm sed}$ [$z$-component (squares), 
$x$-component (triangles)], for colony mean diameter 460 $\mu$m (open symbols), 
and $220$ $\mu$m (solid symbols).
(c) Central region of PDFs of $v_x$ normalized by their standard deviations, for 
$N=11,28,42,128$ and $210$. Dashed black line is a Gaussian fit. 
Full PDFs of (d) $v_x$ and (e) vertical velocity $v_z$.}
\label{figure2}
\end{figure*}

Our simulations of protist suspensions used a model in which the velocity field created by a 
{\it Volvox} is the sum of a downward-pointing Stokeslet and a sourcelet,
\begin{equation} 
{\bf v}({\bf r})=-\frac{3Rv_{\mathrm{sed}}}{4}\!\left[\Bigl(\frac{\hat{z}}{r}\!
+\!\frac{(\hat{z}\!\cdot\!{\bf r}){\bf r}}{r^3}\Bigr)\! +\!\mu R^2 \Bigl(\frac{{\bf n}}{r^3}\! 
-\!\frac{3({\bf n}\!\cdot\!{\bf r}){\bf r}}{r^5}\Bigr)\right],
\nonumber
\end{equation}
where ${\bf n}$ is a unit vector along the colonial axis. The sourcelet represents the near-field flow found by
direct measurements \cite{Drescher10} and in a model with a 
constant force density distributed over the colony surface \cite{Short06},
and is important for the statistics of high fluid velocities. Both 
singularities are cut off at the colony radius. The Stokeslet strength was fixed by an empirical fit
to data on the sedimentation 
velocity as a function of $R$ \cite{waltzing} [$v_{\mathrm{sed}}\simeq \alpha R$, with $\alpha =1$ s$^{-1}$], 
while the relative sourcelet strength $\mu$ was studied in the range $0<\mu<10$. 
We consider the motion of colonies within a non-interacting `ideal gas' model \cite{Locsei}
which, despite its simplicity, gives satisfactory 
agreement with the experiment; the fluid is unbounded, the swimmers are confined to a 
rectangular container (cage) with reflecting walls, and the position ${\bf x}_j$ of the $j$th swimmer and 
its axis vector ${\bf n}_j$ evolve as
\begin{equation}
\dot{\bf x}_j= v_p{\bf n}_j  + {\bf W}_j, \quad  
\dot{\bf n}_j = {\bf \widetilde W}_j~,
\label{eoms}
\end{equation}
where $v_p$ is the propulsion velocity, ${\bf W}_j$ and ${\bf \widetilde W}_j$ are 
white noises with diffusion constants $D$ and $\widetilde{D}$, in 3D and on a 
unit sphere, respectively. They represent the random influences on the motion of {\it Volvox} -- 
irregularities of flagellar beating and, partially, mutual advection of colonies. The latter 
is negligible for the most part, since the typical velocity of the resulting flow is found to be 
much smaller than $v_p$. This is not true, however, when two 
or more colonies come close. Although such events are relatively rare, they are important 
for uniformizing the spatial distribution of {\it Volvox}: without them the bottom-heavy  
colonies would gather in the upper part of the container, contrary to observations.
For the same reason, including the bottom-heaviness and the sedimentation 
into (\ref{eoms}) in the absence of mutual advection would be inconsistent. The 
primary (and minor) consequence of neglecting bottom-heaviness is this model does not reproduce 
the angular distribution of the colonies' axes. Inclusion of 
sedimentation makes only minor changes to the results.

An example of experimental measurements is the time trace of local fluid velocity in
the center of the sample chamber (Fig. \ref{figure1}b).  We see that the observed fluid motion 
is created primarily by the Stokeslets of the swimmers, for when the fluid density was 
increased to match the density of {\it Volvox}, the typical velocity fluctuations were reduced 
drastically.  Yet, the peaks due to the near-field source doublets (from a swimmer passing very 
close to the observation point) remained undiminished. 

On a more quantitative level, we examined the PDF of velocity fluctuations (Fig. \ref{figure2}) 
as a function of the number of colonies in the container, and at various stages in the lifecycle, 
so that the colony size and sedimentation speed vary over a significant range. Data for the smallest 
number of swimmers in the container shows a clear power-law 
tail consistent with the form $v^{-4}$ expected from Eq. \ref{scaling}, and in agreement with
simulations done with pure Stokeslets.  As expected from a gas of Stokeslets, the PDF of the velocity shows 
convergence to a Gaussian with the number of swimmers: for 210 swimmers the Gaussianity persists up 
to 2.5 standard deviations (Fig. \ref{figure2}c), but with clear tails (discussed below).  
Once normalized by the sedimentation 
speed, the standard deviation of the velocity collapses, showing that the fluctuations 
are proportional to the Stokeslet 
strength (Fig. \ref{figure2}b).   In an ideal gas of Stokeslets, the standard deviation of the velocity 
fluctuations grows as $\propto\sqrt{N}$ by virtue of the central limit 
theorem. In the presence of swimmer correlations it should grow faster, but no faster than $\propto N$. 
The observed law lies between these two powers, much closer to the former (Fig. \ref{figure2}b), 
supporting the ideal gas approximation, and distinct from the result $N^{1/3}$ 
found in sedimentation \cite{Segre}, where the mutual advection of particles in each other's 
Stokeslet fields is the sole contribution to velocity fluctuations.  Fluctuations in 
{\it Volvox} suspensions are stronger for larger swimmers, due to their larger sedimentation 
velocity (stronger Stokeslets), and the ratio ${\cal R}\equiv \sigma(v_z)/\sigma(v_x)$ is  
close to $2$ for all $N$. This is found in the numerics with Stokeslets$+$sourcelets 
(whose orientations are uniformly distributed). Without sourcelets, the numerics yield
${\cal R}\sim 2.8\approx 2\sqrt2$. 
For a single Stokeslet offset by a sourcelet, the ratio can be computed analytically, 
averaging over the swimmer's position being replaced by spatial integration, as in (\ref{eq1});  
${\cal R}$ ranges from $1$ (for a randomly directed sourcelet) to $2\sqrt2$ for a 
pure stokeslet.

Inclusion of a sourcelet in the simulations results in tails in the PDFs similar to the experimental 
data (Fig. 2 right). This allows us to conclude that the observed tails are due to the 
near-field component of the swimmers' flow. The tails appear to be exponential, but the range of 
our data is insufficient to prove this. For example, the tail of the data for $N=210$ is 
equally well fit by $P\propto (v^3+Nv_*^3)^{-1}$ with $v_*=25$ $\mu$m/sec.
A similar situation occurs in the PDF of the vertical velocity, where the core convergence to a 
Gaussian is less advanced due to the inherent asymmetry of the Stokeslet field.

\begin{figure}
\begin{center}
\includegraphics[clip=true,width=0.9\columnwidth]{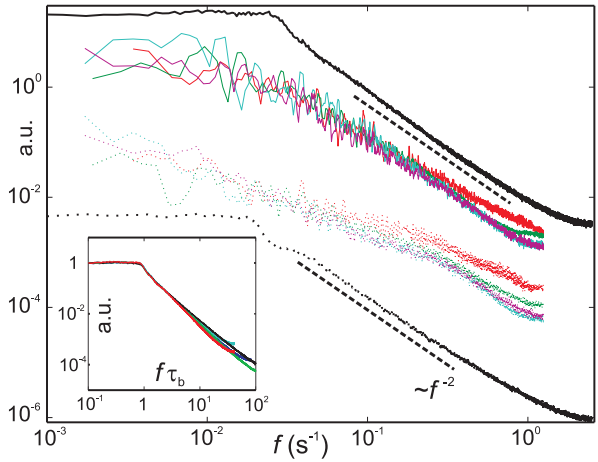}
\end{center}
\caption{(color online). Power-spectra of velocities.   Solid lines and dotted lines are experimental 
spectra of $v_z$ and $v_x$, respectively, for suspensions with mean {\it Volvox} diameter $290$ $\mu$m,
for $N=20$ (red), $70$ (green), $250$ (cyan), and $406$ (magenta). Solid and dotted black lines are  
numerical results for $N=210$ (rescaled in y), with $\mu=4$. Inset: collapse of power-spectra in 
numerics with varying container sizes (0.5-2 cm) and $v_p$ (150-600 $\mu$m/s), with 
$\tau_b$ from $16.7$s to $66.7$ s.}
\label{figure3}
\end{figure}
The velocity power-spectra show a decay close to $f^{-2}$ (Fig. \ref{figure3}), suggesting a Lorentzian power-spectrum of fluctuations $P(f)=(f^2+(2\pi\tau)^{-2})^{-1},$ i.e. an exponential velocity autocorrelation $\langle {\bf v}(0){\bf v}(t)\rangle\propto 
\exp(-t/\tau)$. Supplemented by the Gaussian PDF, this 
equation amounts to modeling the velocity fluctuations as an Orstein-Uhlenbeck stochastic process.
The motion of {\it Volvox} is primarily deterministic. For a concentration $c\sim 100$ cm$^{-3}$ 
the mean free path can be estimated as $(\pi R^2 c)^{-1}\sim 10$ cm, which 
is larger than the container size of $L=1$ cm. Thus, the deterministic term 
in (\ref{eoms}) sets a ballistic time $\tau_b = L/v_p\sim 30$ s which is smaller than the 
diffusive time scale $\tau_d=L^2/D\sim 100$ s or the dephasing time 
$\tau_{ph}=1/\widetilde{D}\sim 100$ s.
Hence, it is $\tau_b$ that sets the correlation time in this ideal gas model. We checked it in the numerics (Fig. \ref{figure3}(inset)), and the characteristic $\tau$ in the experimental data is close to that in the numerics.

In summary, we have introduced a connection between the statistics of velocity fluctuations in
suspensions of swimming protists and the type of force singularity associated with the organism motion.
Experiments and numerical results show clearly the existence of non-Gaussianity in the velocity PDFs,
which are suggested to arise from the details of fluid flow near the organisms.  The greatest challenge 
is a theoretical understanding of the form of the non-Gaussianity, which is known to appear
as well in other contexts, such as inelastic gases \cite{inelastic}.

We are grateful to J.P. Gollub for extensive discussions at an early stage of this work, and
thank K. Drescher, K. Leptos, T.J. Pedley, M. Polin, and I. Tuval for advice, and
D. Page-Croft, J. Milton, and N. Price for technical assistance.  
This work was supported by the Schlumberger Chair fund, the BBSRC, the U.S. DOE, Office of Basic 
Energy Sciences, Division of Materials Science and Engineering, 
Contract DE AC02-06CH11357, and the European Research Council, Grant 247333.

\end{document}